\documentclass[11pt]{article}

\usepackage{fullpage}
\usepackage{latexsym}
\usepackage{amssymb,amsfonts,amsmath,amsthm}

\def\01{\{0,1\}}

\renewcommand{\Pr}{\mbox{\rm Pr}}

\newtheorem{definition}{Definition}
\newtheorem{theorem}{Theorem}

\newtheorem{corollary}[theorem]{Corollary}

\renewcommand{\qed}{\hfill{\rule{2mm}{2mm}}}

\begin{document}
\title{\bf
Connectivity and Structure in Large Networks}

\author{\vspace{1mm} Andr\'as Farag\'o and Rupei Xu  \\
Department of Computer Science \\
The University of Texas at Dallas \\
E-mail: {\tt \{farago, rxx130430\}@utdallas.edu}
}
\date{}
\maketitle

\begin{abstract}
Graph models play a  central role in the description of real life complex networks. 
They aim at constructing graphs that describe the structure of real
systems. The arising graphs, in most cases, are random or random-like, so it is not surprising that there is a large literature on various classes of random graphs and networks. Our key motivating observation is that 
often it is unclear how the strength of
the different models  compare to each  other, e.g., when  will 
a  certain  model  class  contain  another. We are particularly
interested in random  graph models that  arise via (generalized) geometric
constructions. This is  motivated  by  the  fact  that  these graphs can well capture
wireless communication networks. We set up a general framework
to  compare  the  strength  of  random  network models, and present some
results  about  the  equality,  inequality  and  proper containment of
certain model classes.
\end{abstract}

\section{Introduction}
\label{intro}

Large  real  life complex  networks  are  often  modeled by various random graph
contructions, see, e.g. \cite{born,frances,penrosebook} and hundreds of
further references therein. In many cases
it  is  not  at  all  clear  how  the modeling strength of differently
generated random graph  model classes relate  to each other.  We would
like  to  systematically   investigate  such issues. Our
approach was originally motivated to capture properties of the  random
network topology of wireless  communication networks. We started  some
investigations in \cite{farago_alg,farago_jsac,farago_peva}, 
but here we elevate it to a  more general
level  that  makes  it  possible  to compare the strength of
different classes of random network models.

Specifically, we introduce various classes of random graph models that
are significantly more general than the ones that are usually  treated
in the literature, and show relationships among them. One of our main  results
is that no random graph model can fall in the following three  classes
at  the  same  time:  (1)  random  graph  models with bounded expected
degrees;  (2)  random  graph  models  that  are  asymptotically almost
connected; (3) an abstracted version of geometric random graph  models
with two mild restrictions that  we call {\em locality} and  {\em name
invariance}. In other  words, in a  mildly restricted, but  still very
general, class of generalized geometric-style models the  requirements
of bounded  expected degrees  and asymptotic  almost connectivity  are
incompatible.

The rest of the paper is organized as follows.
Section~\ref{classes} introduces the various considered classes of
random network models. Section~\ref{results} presents and proves the
theorems about them.  Section~\ref{appli} shows an application
example, which will (hopefully) convince the reader about the
usefulness of elevating the approach to a higher level of abstraction.
It allows to cut through a lot of complexity that would otherwise
arise in the practically motivated example that is presented here. 
Finally, Section~\ref{concl} concludes the paper
by presenting some open problems.

\section{Classes of Random Graph Models}
\label{classes}

\subsection{General Random Graph Models}

Let us first explain what we mean by random graphs and a random graph
model in the most general sense.
In full  generality, by  a {\em  random graph}  on $n$
vertices
we  mean  a  random  variable that takes its
values in the set of all undirected graphs.
on $n$  vertices. (We use the words {\em vertex} and {\em node}
interchangeably.) Let us denote a random
graph on $n$  nodes by $G_n$.  At this point,  it is still  completely
general,  it  can  be  generated  by  any  mechanism,  with  arbitrary
dependencies among its parts, it is just {\em any} graph-valued random
variable, taking its values among undirected graphs on $n$ nodes.

\begin{definition} {\bf (General random graph model)}
A {\em  random graph  model} is  given by  a sequence  of graph valued
random variables, one for each possible value of $n$:
$${\cal M}=(G_n; \; n\in {\bf N}).$$
The family of all such models is denoted by {\bf GEN}.
\end{definition}

\subsection{Geometric Random Graph Models}

Let us now introduce a model class that reflects a typical feature  of
geometric  random  graph  models.  This  feature  is that in geometric
random graphs the primary random  choice is picking random nodes  from
some  domain  and  then  the  edges  are  already  determined  by some
geometric property  (typically some  kind of  distance) of  the random
nodes. We elevate this approach to an abstract level that, as will  be
shown later, actually turns out to be no less general than the totally
unrestricted model. Our model is built of the following components:

\begin{itemize}

\item {\bf Node variables.} The  nodes are represented by an  infinite
sequence  $X_1,X_2,\ldots$  of  random  variables,  called  {\em  node
variables}. They take their values in an arbitrary (nonempty) set $S$,
which is called the {\em domain} of the model. When a random graph  on
$n$ nodes  is generated,  then we  use the  first $n$  entries of  the
sequence, that is, $X_1,\ldots,X_n$  represent the nodes in  $G_n$. It
is important to note that we  do not require the node variables  to be
independent.

\item {\bf Edge functions.} We denote by $Y_{ij}^{(n)}\in \{0,1\}$ the
indicator
of the edge between nodes $X_i, X_j$ in the random graph $G_n$. Since
loops are not allowed (which is typically the case in geometric
random graph models), we always assume
$i\neq j$, without repeating
this  condition  each  time.  The  (abstract) geometric nature of the
model  is  expressed  by   the  requirement  that  the
random variables $Y_{ij}^{(n)}$
are determined  by  the   nodes  $X_1,\ldots,X_n$,   possibly  with
additional  independent  randomization.  Specifically,  we assume that
there     exist      functions      $f^{(n)}_{ij}$, $1\leq i,j\leq n,$
such that
$$Y_{ij}^{(n)}=f^{(n)}_{ij}(X_1,\ldots,X_n, \xi_{ij})$$ where
$\xi_{ij}$ is
a random variable that is uniformly distributed on $[0,1]$ and is
independent of all the other defining random
variables  of  the  model  (i.e, the  node  variables  and  all  the
other
$\xi_{kl}$ variables). Henceforth the  role of $\xi_{ij}$ is  referred
to as {\em  independent randomization}\footnote{Note that the
specified distribution of $\xi_{ij}$ does not impose a restriction,
since the functions $f^{(n)}_{ij}$ are arbitrary.}. The undirected
nature of
the graph is expressed by the requirement $Y_{ij}^{(n)}=Y_{ji}^{(n)}$,
which can simply be enforced by computing all values for $i<j$ only
and defining the $i>j$ case by exchanging $i$ and $j$.

\end{itemize}

We use the following {\bf notational convention:} whenever a function
is
distinguished by certain parameters within some family of functions,
such as $f^{(n)}_{ij}$ above, then it is assumed that the function
``knows" its own parameters. In other words, the parameter values can
be used in the definition of the function. Conversely, whatever
information is used in computing the function should occur either as a
variable or an explicitely shown parameter.

\begin{definition}
{\bf (Abstract geometric model)}
The class of all models that have the structure explained above
is called {\bf GEOM}.
\end{definition}

A model ${\cal  M}\in {\bf GEOM}$,  no matter how  general it can  be,
still has a restricted structure. Therefore, one may ask whether {\em
every} model in {\bf GEN} can be represented in such a way. To make it
precise when two models or model classes are considered equivalent,
let us introduce the following definition.

\begin{definition}
{\bf (Equivalence)}
Two random graph models
${\cal M}=(G_n;  \; n\in {\bf N})$ and
$\widetilde{\cal M}=(\widetilde G_n;  \; n\in {\bf N})$
are called {\em equivalent}, denoted by ${\cal M}\sim \widetilde{\cal
M}$,  if for any graph
$G$   on $n$ vertices
$$\Pr(G_n=G)=\Pr(\widetilde G_n=G)$$
holds,
where equality of graphs means that they are isomorphic.
\end{definition}

\begin{definition} \label{wedge}
{\bf (Containment, equivalence, interesection and  disjointness of
model classes)}
Let ${\bf C_1,C_2}$  be two classes of random graph models. We say
that ${\bf C_2}$ contains ${\bf C_1}$, denoted by
${\bf C_1}\preceq {\bf C_2}$,
if for every
${\cal M}_1\in {\bf C_1}$ there is an
${\cal M}_2\in {\bf C_2}$, such that
${\cal M}_1\sim {\cal M}_2$.
If
${\bf C_1}\preceq {\bf C_2}$ and
${\bf C_2}\preceq {\bf C_1}$ both hold, then the two classes are
called {\em equivalent}, denoted by
${\bf C_1}\simeq {\bf C_2}$.
The {\em intersection} of ${\bf C_1}$ and ${\bf C_2}$, denoted by
${\bf C_1}\wedge {\bf C_2}$, is the set of models ${\cal M}$ with
the property that there exist models
${\cal M}_1\in {\bf C_1}$ and
${\cal M}_2\in {\bf C_2}$, such that
${\cal M}\sim {\cal M}_1$ and ${\cal M}\sim {\cal M}_2$.
If no model $\cal M$ has this property,
then the classes ${\bf C_1,C_2}$ are called {\em
disjoint}.
\end{definition}

Now we may ask whether
${\bf
GEOM}\simeq {\bf GEN}$ holds or not. We  show later that it does,
even
with more restrictions on {\bf  GEOM}. To this end, we  introduce some
restricting conditions  to the  model class  {\bf GEOM}.  As a  simple
notation, whenever some restrictions $R_1,\ldots,R_k$ are applied, the
arising class is denoted by ${\bf GEOM}(R_1,\ldots,R_k)$.

\subsection{Subclasses of {\bf GEOM}}

The first considered restriction is called {\em locality}.
Up to now we allowed that an  edge in $G_n$ can  depend on all
the  nodes,  and  the  dependence  expressed  by  the   $f^{(n)}_{ij}$
functions can be arbitrary and different  for each edge. To  get a
little closer to
the usual geometric random graph model (see, e.g.,
\cite{penrosebook}),
we introduce the condition of locality. Informally, it restricts the
dependence
of an edge to its endpoints, in a homogeneous way, but still via an
{\em arbitary} function.

\begin{definition}\label{deflocal} {\bf  (Locality)} A  model ${\cal  M}\in
{\bf GEOM}$ is called {\em local}, if for every $n$
and $i,j\leq n$  the existence of  an edge between  $X_i, X_j$ depends
only on these  nodes. Moreover, the  dependence is the  same for every
$i,j$, possibly  with independent  randomization. That  is, there  are
functions $f^{(n)}$ such that  the edge indicators are  expressible as
$$Y_{ij}^{(n)}=f^{(n)}(X_i,X_j,    \xi_{ij})$$    where     $\xi_{ij}$
represents the independent randomization. The set of local models in
{\bf GEOM} is denoted by ${\bf GEOM}(loc)$.
\end{definition}
{\em Note:} with
our notational convention $f^{(n)}$ can depend on its variables and on
$n$. On the other hand, it has no access to the value of $i$ and  $j$,
unless they are somehow contained in  $X_i, X_j$, in a way that  makes
it  possible  to  extract  them  without  using anything else than the
explicitly listed information.

Another restriction that we consider is a condition on the
distribution of the vertices. To introduce it,
let us first recall a concept from probability theory, called
exchangeability.

\begin{definition}\label{defex}{\bf  (Exchangeable  random  variables)}  A
finite sequence $\xi_1,\ldots,\xi_n$  of random variables  is called
{\em  exchangeable}  if  for any permutation
$\sigma$   of    $\{1,\ldots,n\}$,   the    joint   distribution    of
$\xi_{1},\ldots,\xi_{n}$  is  the  same  as  the joint distribution of
$\xi_{\sigma(1)},\ldots,\xi_{\sigma(n)}$. An infinite sequence of
random variables is called exchangeable if every finite initial
segment of the sequence is exchangeable.
\end{definition}

Exchangeability  can  be  equivalently  defined  such  that when
taking any
$k\geq 1$ of  the random variables,  say,
$\xi_{j_1},\ldots,\xi_{j_k}$,
their joint distribution  does not depend  on which particular  $k$ of
them  are  taken, and in which order.  Note  that  independent,  identically   distributed
(i.i.d.) random variables are always exchangeable, but the converse is
not true, so this is a larger family.

Now let us introduce the condition that we use to restrict the
arbitrary dependence of node variables.

\begin{definition}\label{defname}{\bf  (Name invariance)}
A
random graph model ${\cal M}\in {\bf GEOM}$ is called {\em name
invariant}, if its node variables are exchangeable.
The class of such models is denoted by ${\bf GEOM}(inv)$.
\end{definition}

We call it the {\em name invariance} of the model because it means the
names (the indices) of the nodes are irrelevant in the sense that  the
joint probabilistic behavior of any fixed number of nodes is invariant
to renaming  (reindexing) the  nodes. In  particular, it  also implies
that  each  single  node  variable  $X_i$  has  the  same  probability
distribution  (but  they  do  not  have  to  be  independent).

A  simple  example  for  a  dependent,  yet still name invariant, node
generation process is  a ``clustered uniform"  node generation. As  an
example,  let  $S$  be  a  a  sphere in 3-dimensional space, i.e., the
surface of a 3-dimensional  ball. Let $R$ be  the radius of the  ball.
Let us first generate a pivot point $Y$ uniformly at random from  $S$.
Then  generate  the  nodes  $X_1,X_2,\ldots$  uniformly  at random and
independently of each other from  the neighborhood of radius $r\ll  R$
of the  random pivot  point $Y$  (within the  sphere). It is directly
implied by the construction  that exhangeability holds. Moreover,  any
particular $X_i$ will be  uniformly distributed over the  {\em entire}
sphere, since $Y$ is uniform over  the sphere. On the other hand,  the
$X_i$  are  far  from  independent  of  each other, since they cluster
around $Y$, forcing any  two of them to  be within distance $2r$.  The
example  can  be  generalized  to  applying  several  pivot points and
non-uniform distributions, creating a more sophisticated clustering.

It is worth mentioning that {\em any} finite  sequence
$X_{1},\ldots,X_{n}$  of random variables can be easily transformed
into an exchangeable sequence by taking a {\em random} permutation
$\sigma$  of  $\{1,\ldots,n\}$ and defining the  transformed sequence
by $\widetilde X_i=X_{\sigma(i)}.$ The resulting joint distribution
will be
$$\Pr(\widetilde X_1=x_1, \ldots, \widetilde X_n=x_n)=
\frac{1}{n!}
\sum_\sigma\Pr(X_{\sigma(1)}=x_1, \ldots,X_{\sigma(n)}=x_n)
$$
where $\sigma$ in the summation runs over all possible permutations
of  $\{1,\ldots,n\}$.
Even though this simple construction does not work for infinite
sequences, in many practically relevant cases there is
vanishing
difference between a very long finite and an actually infinite
sequence.

A stronger restriction is if we want the node variables to be
independent, not just exchangeable.

\begin{definition}{\bf  (Free geometric model)}
A
random graph model ${\cal M}\in {\bf GEOM}$ is called {\em free},
if its node variables are mutually independent.
The class of such models is denoted by ${\bf GEOM}(free)$.
\end{definition}

\subsection{Other Model Classes}

We define some other classes of random graph models, relating to some
properties that are important in the applications of these models.

\begin{definition}{\bf  (Bounded  expected  degree model)} A
random graph model  ${\cal M}\in {\bf  GEN}$ is called  a {\em bounded
expected  degree  model}  if  there  exists  a  constant $C$ such that
$$\overline  d(n)=\frac{2{\rm  E}(e(G_n))}{n}\leq  C$$  for every $n$,
where $e(G_n)$  denotes the  number of  edges in  $G_n$, and {\rm E}
stands for the expected value. The class
of bounded expected degree models is denoted by {\bf BD}. \end{definition}

Since $2e(G_n)/n$ is the average degree in $G_n$, therefore, $\overline
d(n) = 2{\rm E}(e(G_n))/n$ is  the expected average degree. 
It can be interpreted as the expected degree of a randomly chosen node. Often  the
expected degree of  each individual node  is also equal  to $\overline
d(n)$, but in a general model it  may not hold. Note that even if  the
expected degree of each node is equal to the expected average  degree,
it does not mean that the  actual (random) degrees are also equal,  so
$G_n$ may be far from regular.

Another important property of random graph models is asymptotically
almost sure (a.a.s.) connectivity.

\begin{definition}  {\bf  (Connected  model)}  A  random graph model ${\cal
M}=(G_n; \; n\in {\bf N})\in  {\bf GEN}$ is called {\em  connected} if
$$\lim_{n\rightarrow\infty}\Pr(\mbox{\rm $G_n$ is connected})=1.$$ The
class of connected models is denoted by {\bf CONN}. \end{definition}

Often the requirement of full connectivity is too strong, so we define
a relaxed version of it and the corresponding model class.

\begin{definition}\label{defbetaconn}
{\bf ($\beta$-connectivity)}
For a real number $0\leq \beta\leq 1$, a graph $G$ on $n$ vertices is
called
$\beta$-connected if $G$ contains a connected component on at least
$\beta n$ nodes.
\end{definition}

When we consider  a sequence of  graphs with different  values of $n$,
then the parameter $\beta$ may depend  on $n$. When this is the  case,
we write $\beta_n$-connectivity. Note that even if $\beta_n\rightarrow
1$, this is still weaker then full connectivity in the limit. For
example, if $\beta_n=1-1/\sqrt{n}$, then we have $\beta_n\rightarrow
1$, but there can be still $n-\beta_nn=\sqrt{n}$
nodes that are not part of the largest connected component.

\begin{definition} {\bf  ($\beta_n$-connected model)}
A
random graph model ${\cal M}=(G_n;  \; n\in {\bf N})\in {\bf  GEN}$ is
called             $\beta_n$-{\em             connected}            if
$$\lim_{n\rightarrow\infty}\Pr(\mbox{\rm           $G_n$            is
$\beta_n$-connected})=1.$$ The class  of $\beta_n$-connected models
is denoted by $\beta_n$-{\bf CONN}. \end{definition}

It is  clear from  the definitions  that with  $\beta_n\equiv 1$,  the
class 1-{\bf CONN} is the same as {\bf CONN}. But if we only know that
$\beta_n\rightarrow  1$,  then  $\beta_n$-{\bf  CONN} becomes a larger
class.

Finally,  let  us  define  some  classes that restrict the indepedence
structure of the edges. Let $e$ be a (potential) edge. We regard it as
a 0-1 valued  random variable, indicating  whether the edge  is in the
random  graph  or  not.  The  probability  that  an edge $e$ exists is
$\Pr(e=1)$, but we  simply denote it  by $\Pr(e)$. We  similarly write
$\Pr(e_1,\ldots,e_k)$ instead of $\Pr(e_1=1,\ldots,e_k=1)$.

\begin{definition}  {\bf  (Independent disjoint edges) }  A
random graph model
${\cal
M}=(G_n; \; n\in {\bf N})\in {\bf GEN}$ is said to have {\em
independent disjoint edges}
if any set $e_1\ldots,e_k$ of pairwise disjoint edges are independent
as random variables.
That is,
$$\Pr(e_1,\ldots,e_k)=\Pr(e_1)\ldots\Pr(e_k)$$
holds whenever $e_1,\ldots,e_k$ are pairwise disjoint.
The  class of  models with
independent disjoint edges
  is denoted
by {\bf IDE}. \end{definition}

\begin{definition}  {\bf  (Positively correlated edges) }  A
random graph model
${\cal
M}=(G_n; \; n\in {\bf N})\in {\bf GEN}$ is said to have {\em
positively correlated edges}
if any set $e_1\ldots,e_k$ of distinct edges are positively correlated
in the sense of
$$\Pr(e_1,\ldots,e_k)\geq \Pr(e_1)\ldots\Pr(e_k).$$
The  class of  models with
positively correlated edges
  is denoted
by {\bf POS}. \end{definition}

\section{Results}
\label{results}

Let  us  first  address  the  question  how  the  various restrictions
influence the modeling strength of {\bf GEOM}. The motivation is  that
one might  think that  a concept  like locality  imposes a significant
restriction on the model. After all, it severely restricts which  node
variables  can  directly  influence  the  existence  of  an  edge. For
example, it seems to exclude situations when the existence of an  edge
between $X_i$ and $X_j$ is based  on whether one of them is  among the
$k$  nearest  neighbors  of  the  other,  according  to  some distance
function (often called $k$-nearest neighbor graph).

Surprisingly, it turns  out that locality  alone does not  impose {\em
any} restriction at all on the generality of the model. Not just
any model in {\bf GEOM} can be expressed by a
local one, but this remains true even if we want to express an {\em
arbitrary
} random graph model in {\bf GEN}.

\begin{theorem}\label{thm1}
Let
$\widetilde{\cal M}=(\widetilde G_n;  \; n\in {\bf N})\in {\bf GEN}$
be an arbitrary random graph model. Then there exists
another model
${\cal M}=(G_n;  \; n\in {\bf N})\in {\bf GEOM}(loc)$
such that ${\cal M}\sim \widetilde{\cal M}$.
\end{theorem}

\noindent {\bf Proof.}
Let $\widetilde Y_{ij}^{(n)}$ denote the edge indicators in
$\widetilde{\cal M}$. We show that a
${\cal M}\in {\bf GEOM}(loc)$ can be chosen
such that its edge indicators $Y_{ij}^{(n)}$ satisfy
$Y_{ij}^{(n)} = \widetilde Y_{ij}^{(n)}$, which implies that the two
models are equivalent.

Let $Q$ be the set of all 0-1 matrices of all possible finite
dimensions. For the domain $S$ of $\cal M$ we choose the set of all
infinite sequences with entries in $Q$.
  Let us define the node variable $X_i$ such that
$X_i = (Z_i^{(1)},Z_i^{(2)},\ldots)$,
where $Z_i^{(n)}$ is an $(n+1)\times n$ sized 0-1 matrix with entries
$Z_i^{(n)}[k,\ell]=\widetilde Y_{k,\ell}^{(n)}$ for $k\neq \ell$ and
$k,\ell\leq n$,
$Z_i^{(n)}[k,k]=0$ and the last row $Z_i^{(n)}[n+1,\,.\,]$ contains
the binary encoding of $i$. Then the edge functions for
$\cal M$ can be defined as $$f^{(n)}(X_i,X_j,
\xi_{ij})=Z_i^{(n)}[i,j].$$
This indeed defines $f^{(n)}$, since knowing $n$ the
matrix $Z_i^{(n)}$ can be obtained as the $n^{th}$ component of $X_i$.
The value
of $i$ can be read out from the last row of $Z_i^{(n)}$.
Similarly, the value of $j$ can be read out from the last row of
$Z_j^{(n)}$, which is the $n^{th}$ component of $X_j$.
Then the value
of $Z_i^{(n)}[i,j]$ can be looked up.
(The functions do not use the independent randomization). This
definition directly implies that $\cal M$ is local, as  $f^{(n)}$
does not use node variables other than $X_i,X_j$ and the same
function applies to any pair of nodes. Furthermore,
$$Y_{ij}^{(n)} =
f^{(n)}(X_i,X_j,  \xi_{ij})=Z_i^{(n)}[i,j]=
\widetilde Y_{ij}^{(n)}$$ holds, completing the proof.

\hfill $\spadesuit$

Next we show that a similar result holds for the restriction of name
invariance.

\begin{theorem}\label{thm2}
Let
$\widetilde{\cal M}=(\widetilde G_n;  \; n\in {\bf N})\in {\bf GEN}$
be an arbitrary random graph model. Then there exists a
another model
${\cal M}=(G_n;  \; n\in {\bf N})\in {\bf GEOM}(inv)$
such that ${\cal M}\sim \widetilde{\cal M}$.
\end{theorem}

\noindent {\bf Proof.}
We show that the
name invariant model
${\cal M}\in {\bf GEOM}(inv)$ can be
chosen such that its edge indicators $Y_{ij}^{(n)}$ satisfy
$Y_{ij}^{(n)} = \widetilde Y_{ij}^{(n)}$,
where the $\widetilde Y_{ij}^{(n)}$ denote the edge indicators in
$\widetilde{\cal M}$.

Let $Z_n=[\widetilde Y_{ij}^{(n)}]$ be an $n\times n$  matrix,
containing
all edge indicators of $\widetilde G_n$. Define $X_i$ as an infinite
sequence
$$X_i=(Z_1,Z_2,\ldots).$$
Since $X_i$ is defined without using the value of $i$, we have
that all the $X_i$ are equal, which is a trivial case of name
invariance.
(All
random node variables being equal, re-indexing clearly cannot change
anything.) Then, following the edge function format of
{\bf GEOM},
we can define the edge functions by
$$  f^{(n)}_{ij}(X_1,\ldots,X_n, \xi_{ij}) = Z_n[i,j].$$
(The independent randomization is not used.)
This edge function is well defined, since, knowing $n$, the array
$Z_n$ can be read out from any of the $X_i$ and in the general
{\bf GEOM} model
the functions can directly depend on $i$ and $j$. As, by definition,
$Z_n[i,j]=\widetilde Y_{ij}^{(n)}$, we obtain
$$Y_{ij}^{(n)} =
 f^{(n)}_{ij}(X_1,\ldots,X_n, \xi_{ij}) = Z_n[i,j]=
\widetilde Y_{ij}^{(n)}$$
which completes the proof.

\hfill $\spadesuit$

\medskip
Since we know by definition
${\bf GEOM}(loc) \preceq {\bf GEOM}$ and
${\bf GEOM}(inv) \preceq {\bf GEOM}$, as well as
${\bf GEOM} \preceq {\bf GEN}$,
the theorems immediately imply the following corollary.

\begin{corollary} \label{cor1}
${\bf GEOM}(loc) \simeq
{\bf GEOM}(inv) \simeq
{\bf GEOM}\simeq {\bf GEN}$.
\end{corollary}

We have seen above  that neither locality  nor name invariance  can
restrict
full generality. Both restrictions, if applied alone, still allow that
an {\em arbitrary} random graph model is generated.
This situation naturally  leads to the  question: what happens  if the
two restrictions are applied {\em together?} At first, one might think
about it  this way:  if the  set of  local models  and the set of name
invariant models  are both  equal to  the set  of general models, then
their intersection should also be the same. This would mean that  even
those models that  are both local  and name invariant  are still fully
general.

The above  argument, however,  is not  correct. Although
Corollary~\ref{cor1} implies
$${\bf GEOM}(loc) \wedge {\bf GEOM}(inv) \simeq {\bf GEN}$$ (see
Definition~\ref{wedge} for the $\wedge$ operation),
it does not imply that
$${\bf GEOM}(loc, inv) \simeq {\bf GEOM}(loc) \wedge {\bf GEOM}(inv)$$
also holds. In fact,  the latter does not hold, which will be obtained
as a consequence
of the following theorem. The theorem proves the surprising fact
that joint locality and name invariance, without any further
restriction,  makes it impossible that
a model satisfies bounded expected degree and (almost) connectivity
at the same time.

\begin{theorem} \label{thm3} Let $\beta_n\rightarrow 1$ be a
sequence of positive reals. Then
$${\bf BD}\,\wedge \,\beta_n{\bf {\rm -}CONN}\,\wedge\,
{\bf GEOM}(loc,inv)=\emptyset$$
holds.
\end{theorem}

\noindent {\bf Proof.}
Consider a model
${\cal M}=(G_n;  \; n\in {\bf N})\in {\bf GEOM}(loc, inv)$.
Let $I_n$ denote the (random) number of isolated nodes in
$G_n$. First we show that
\begin{equation}\label{lb}
{\rm E}(I_n) \geq n\left(1-\frac{\overline d(n)}{n-1}\right)^{n-1}
\end{equation}
holds\footnote{
It is worth noting that even when
${\rm E}(I_n)\rightarrow\infty$ is the case, this fact alone
may not {\em a priori} preclude the possibility of
a.a.s. $\beta_n$-connectivity, even with $\beta_n\equiv 1$.
For example, if
$G_n$ is connected with probability $1-1/\sqrt{n}$ and consists of
$n$ isolated nodes with probability $1/\sqrt{n}$, then
${\rm E}(I_n)=n/\sqrt{n}\rightarrow\infty$, but
$\Pr(\mbox{\rm $G_n$ is connected})=1-1/\sqrt{n}\rightarrow 1.$}.
Note that since our model is abstract and does not involve
any real geometry, one has to be careful to avoid using such
intuition that may appeal geometrically,
but does not follow from the abstract model.

First, observe the following: name invariance implies that
for any function
$g$ of the node variables and for any permutation $\sigma$ of
$\{1,\ldots,n\}$ we have
$${\rm E}(g(X_1,\ldots,X_n))=
{\rm E}(g(X_{\sigma(1)},\ldots,X_{\sigma(n)})).$$
Since the probability that a particular node has any given
degree
$k$ is also expressible by such a function, therefore, the probability
distribution of the node degree must be the same for all
nodes
(but the degrees, as random variables, may not be independent).
As a consequence, the expected degree
of each
node is the same, which then must be equal to the
expected average degree $\overline d(G_n)$.

Let us pick a node $X_i$.  We derive a lower bound on  the probability
that  $X_i$  is  isolated,  i.e.,  that its  degree  is 0. Due to the above
symmetry considerations, it does not  matter which node is chosen,  so
we can take $i=1$.  Let ${\cal I}_n$ be  the (random) set of  isolated
nodes  in  $G_n$.  What  we  want  to  compute  is  a  lower  bound on
$\Pr(X_1\in {\cal I}_n)$. Then we are going to use the fact that
$${\rm E}(I_n)=
{\rm E}(|{\cal I}_n|)= \sum_{i=1}^n \Pr(X_i\in {\cal I}_n)
$$
Note that, due to the linearity of expectation, this remains true even
if the  events $\{X_i\in {\cal I}_n\}$ are not independent, which is
typically the case. Then, by the symmetry considerations, we can
utilize that $\Pr(X_i\in {\cal I}_n)$ is independent of $i$, yielding
${\rm E}(I_n)=  n\Pr(X_1\in {\cal I}_n). $

In order to derive a lower bound on $\Pr(X_1\in {\cal I}_n)$,
we need a fundamental result from probability theory,
called {\em de Finetti's Theorem}\footnote{It was first published in
\cite{definetti}. Being a classical result, it can be found in many
advanced textbooks on probability.}.
 This theorem says
that if an infinite sequence $\xi_1,\xi_2,\ldots$ of 0-1 valued random
variables\footnote{Various  extensions exist to more general
cases, see, e.g., \cite{kallenberg}, but for our purposes the simplest
0-1
valued case is sufficient.} is exchangeable, then the following hold:
\begin{description}
\item[{\rm (i)}]
The limit
\begin{equation}\label{eta}
\eta= \lim_{N\rightarrow\infty}\frac{\xi_1+\ldots+\xi_N}{N}
\end{equation}
exists\footnote{
Note that exhangeability implies that all $\xi_i$ have the same
expected value, so in case they were independent,
then the strong law of large numbers would apply and the limit would
be the common expected value, with probability 1.  Since, however, the
$\xi_i$ are not assumed independent   (only
exchangeable), therefore, the average may not tend to a constant, it
can be a non-constant random variable in $[0,1]$.}
with probability 1.

\item[{\rm (ii)}]
For any $N$ and for any system $a_1, \ldots,a_N\in \{0,1\}$ of
outcomes with
$s=\sum_{i=1}^N a_i$
$$\Pr(\xi_1=a_1,\ldots,\xi_N=a_N)=\int_0^1 x^s(1-x)^{N-s}dF_\eta(x)$$
holds, where $F_\eta$ is the probability distribution function of
$\eta$.

\item[{\rm (iii)}]
The $\xi_i$ are
conditionally independent
and identically distributed (conditionally i.i.d.), given $\eta$, that
is,
$$\Pr(\xi_1=a_1,\ldots,\xi_N=a_n\,|\,\eta)=
\prod_{i=1}^N \Pr(\xi_i=a_i\,|\,\eta).
$$

\end{description}

Informally, de  Finetti's theorem  says that  exchangeable 0-1  valued
random variables,  even if  they are  not independent,  can always  be
represented as a mixture of Bernoulli systems of random variables.  It
is important to  note, however, that  even though the  statements (ii)
and  (iii)  refer   to  finite  initial   segments  of  the   sequence
$\xi_1,\xi_2,\ldots,$ it is necessary  that the entire {\em  infinite}
sequence is  exchangeable. For  finite sequences  the theorem  may not
hold, counterexamples are known for the finite case \cite{stoyanov}.

Let us now define the infinite sequence of 0-1 valued random variables
$$e_{j}=f^{(n)}(X_1,X_j, \xi_{1j}), \;\;\;\;\; j=2,3\ldots$$ Of these,
$e_2,\ldots,e_n$ are the indicators of the edges with one endpoint  at
$X_1$. But  the function  $f^{(n)}$ is  defined for  any $(x,y,  z)\in
S\times S\times  [0,1]$, so  nothing prevents  us to  define the  {\em
infinite} sequence $e_j; \, j=2,3,\ldots$, by taking more  independent
and uniform $\xi_{1j}\in [0,1]$ random variables.

Observe now that  the sequence $e_j;  \, j=2,3,\ldots$ is  an infinite
exchangeable  sequence  of  0-1  valued  random  variables.  Only  the
exchangeability   needs   proof.   If   we   take   any   $k$  indices
$j_1,\ldots,j_k$,  then  the  joint  distribution  of $e_{j_1},\ldots,
e_{j_k}$  depends  only on  the  joint  distribution  of
$X_{j_1},\ldots,
X_{j_k}$,   plus   the   independent   randomization.  If  we  replace
$j_1,\ldots,j_k$ by  other $k$  indices, then  it will  not change the
joint distribution  of the  $k$ node  variables, due  to their assumed
exhangeability. The independent randomization also does not change the
joint distribution,  since the  $\xi_{1j}$ are  i.i.d, so  it does not
matter which  $k$ are  taken. Furthermore,  the locality  of the model
implies that each $e_j$ depends on one $X_j$ (besides $X_1$) so taking
another $k$ cannot change how  many node variables will any  subset of
the  $e_j$  share.  Thus,  for  any  $k$,  the  joint  distribution of
$e_{j_1},\ldots, e_{j_k}$  does not  depend on  which $k$  indices are
chosen,  proving   that  $e_j;   \,  j=2,3,\ldots$   is  an   infinite
exchangeable sequence of 0-1 valued random variables.

Now, by de Finetti's Theorem, there is a random variable $\eta\in
[0,1]$, such that the $e_j$ are conditionally i.i.d, given $\eta$.
Then we can write
\begin{eqnarray}      \nonumber
\Pr(X_1\in {\cal I}_n) & = & \Pr(e_2=\ldots=e_n=0) \\ \nonumber
& = & {\rm E} (\Pr(e_2=\ldots=e_n=0\,|\,\eta)) \\ \nonumber
& = & {\rm E} \left(\prod_{j=2}^n(\Pr(e_j=0\,|\,\eta))\right) \\
& = & {\rm E} \left(\prod_{j=2}^n(1-\Pr(e_j=1\,|\,\eta))\right). \label{cidd}
\end{eqnarray}
Notice that $\Pr(e_j=1\,|\,\eta)$ is the probability that an edge
exists between $X_1$ and $X_j$, conditioned on $\eta$. Consequently,
$\xi=\Pr(e_j=1\,|\,\eta)$ is a random variable,
depending on $\eta$.
 At the same time, it does not
depend on
$j$, as by
de Finetti's theorem, the $e_j$ are conditionally i.i.d, given $\eta$,
so it does not matter which $j$ is taken in $\xi=\Pr(e_j=1\,|\,\eta)$.
Thus, we can continue (\ref{cidd}) as
\begin{equation}\label{cidd2}
\Pr(X_1\in {\cal I}_n) =
{\rm E} \left(\prod_{j=2}^n(1-\xi)\right)=
{\rm E} \left((1-\xi)^{n-1}\right).
\end{equation}
We can now observe that $\xi\in [0,1]$ and the function $g(x)=(1-x)^n$
is convex in $[0,1]$, so we may apply Jensen's inequality.
Jensen's well known inequality says that for any
random
variable $\zeta$ and for any convex function $g$ the inequality
${\rm E}\big(g(\zeta)\big)\geq
 g\big({\rm E}(\zeta)\big)$ holds, which
 is a consequence of the definition of convexity. Thus, we can further
continue (\ref{cidd2}), obtaining
$$
\Pr(X_1\in {\cal I}_n) =
{\rm E} \left((1-\xi)^{n-1}\right)\geq
\left(1-{\rm E}(\xi)\right)^{n-1}.
$$
Note that
${\rm E}(\xi)={\rm E}(\Pr(e_j=1\,|\,\eta))=\Pr(e_j=1)$ is the
probability that an edge exists
between $X_1$ and $X_j$. By name invariance, this is the same
probability for any two nodes, let $p_n$ denote this common value.
Thus,
$$
\Pr(X_1\in {\cal I}_n) \geq (1-p_n)^{n-1}$$
follows. We know that there are $n-1$ potential edges adjacent to each
node, each with probabilty $p_n$. Therefore,
despite the
possible dependence of edges, the linearity of expectation implies the
expected degree of each node under our conditions is $(n-1)p_n$, which
is also equal to $\overline d(n)$. We can then substitute
$p_n=\overline d(n)/(n-1)$, which yields
$$
\Pr(X_1\in {\cal I}_n)
\geq \left(1-\frac{\overline d(n)}{n-1}\right)^{n-1},
$$
implying
$$
{\rm E}(I_n) =
n\Pr(X_1\in {\cal I}_n)
\geq n\left(1-\frac{\overline d(n)}{n-1}\right)^{n-1}.
$$
Assume now there is a model ${\cal M}'\in {\bf BD}$ with
${\cal M}'\sim{\cal M}.$ This means,
there is a constant
$C$ with
$\overline d(n)\leq C$ for every $n$. Then
$$
\left(1-\frac{\overline d(n)}{n-1}\right)^{n-1}\geq
\left(1-\frac{C}{n-1}\right)^{n-1}
\rightarrow {\rm e}^{-C},$$
so there exist constants $a>0$ and $n_0\in {\bf N}$, such that
${\rm E}(I_n) \geq an$ holds for every $n\geq n_0$.

Now take a sequence $\beta_n\in [0,1]$ with $\beta_n\rightarrow 1$.
We are going to show that the probability
$\Pr(\mbox{\rm $G_n$ is $\beta_n$-connected})$ cannot tend to 1,
meaning that for any model ${\cal M}''$ with
${\cal M}''\sim{\cal M}$ it holds that
${\cal M}''\notin \beta_n{\bf -CONN}$.

Set  $s_n=  \Pr(I_n\leq  (1-\beta_n)n)$.  Then $\Pr(\mbox{\rm $G_n$ is
$\beta_n$-connected})\leq s_n$ must hold, since $\beta_n$-connectivity
implies  that  there  may  be  at  most $(1-\beta_n)n$ isolated nodes.
Consider now the random  variable $\gamma_n=n-I_n$. The definition  of
$\gamma_n$  implies  $\gamma_n\geq  0$  and  ${\rm E}(\gamma_n)=n-{\rm
E}(I_n)$. Therefore, ${\rm E}(\gamma_n)\leq (1-a)n$ holds for $n\geq
n_0$. Moreover, the definition also directly implies that the events
$\{I_n\leq (1-\beta_n)n\}$ and $\{\gamma_n\geq \beta_nn\}$ are
equivalent. Thus, we can write, using Markov's
inequality for nonnegative random variables:
$$s_n=\Pr(I_n\leq  (1-\beta_n)n)=
\Pr(\gamma_n\geq  \beta_nn)\leq
\frac{{\rm E}(\gamma_n)}{\beta_nn} \leq \frac{(1-a)n}{\beta_nn}=
\frac{1-a}{\beta_n}.
$$
Since we know that $a>0$ is a constant and $\beta_n\rightarrow 1$,
therefore, there must exist a constant $b<1$, such that $s_n\leq b$
holds for all large enough $n$. This, together with
$\Pr(\mbox{\rm $G_n$ is $\beta_n$-connected})\leq s_n$,
proves that the assumptions we made, that is,
${\cal M}\in {\bf GEOM}(loc, inv)$ and
${\cal M}\sim{\cal M}'\in \beta_n{\bf -CONN}$, together imply
that there is no
${\cal M}''\sim{\cal M}$ with
${\cal M}''\in {\bf BD}$, proving the theorem.

\hfill $\spadesuit$

\medskip

As a corollary, we obtain that ${\bf GEOM}(loc, inv)$ is smaller
than ${\bf GEOM}(loc)$ and ${\bf GEOM}(inv)$.

\begin{corollary} \label{cor2}
${\bf GEOM}(loc, inv) \not\simeq {\bf GEOM}(loc)$ and
${\bf GEOM}(loc, inv) \not\simeq {\bf GEOM}(inv)$.
\end{corollary}

\noindent {\bf Proof.}
Let
${\cal M}=(G_n;  \; n\in {\bf N})$ be a model in which $G_n$ is chosen
unformly at random from the set of all connected graphs with maximum
degree at most 3. It follows from this construction that
${\cal M}\in {\bf BD}\,\wedge \,{\bf CONN}$, implying
${\cal M}\in {\bf BD}\,\wedge \,\beta_n{\bf {\rm -}CONN}$ for any
$\beta_n$. Then
Theorem~\ref{thm3} implies ${\cal M}\notin {\bf GEOM}(loc, inv)$.
Since, naturally, ${\cal M}\in {\bf GEN}$, therefore, it follows that
${\bf GEOM}(loc, inv) \not\simeq {\bf GEN}$.
As we know from Corollary~\ref{cor1} that
${\bf GEOM}(loc) \simeq
{\bf GEOM}(inv) \simeq
{\bf GEN}$, we obtain
${\bf GEOM}(loc, inv) \not\simeq {\bf GEOM}(loc)$ and
${\bf GEOM}(loc, inv) \not\simeq {\bf GEOM}(inv)$.

\hfill $\spadesuit$

\section{An  Application}
\label{appli}

In this application example we model a mobile wireless ad hoc network,
that is, a network in  which wireless nodes communicate to  each other
directly, without a supporting infrastructure. The initial position of
each node is  chosen in the  following way. Let  $P$ be a  probability
measure over  a planar  domain $D$.  First we  choose $k$ pivot points
independently at random, using $P$. Then the actual node positions are
generated such  that each  potential node  is chosen  independently at
random from $P$, but it is kept only if it is within a given  distance
$d_0$ to  at least  one of  the random  pivot points,  otherwise it is
discarded.  Note  that  this  way  of  generating the nodes makes them
dependent, as the non-discarded  ones cluster around the  random pivot
points, thus modeling a clustered, non-independent node distribution.

The mobility of the nodes in this example is modeled in the  following
way. Over some time horizon $T_n$, that may depend on $n$, the  number
of  nodes,  each  node  moves  along  a  random curve from its initial
position with a constant speed $v_0$.  The curve is chosen from a  set
$\cal  C$  of  available  potential trajectories  in  $D$. For
simplicity, it is
assumed that each  curve can be  identified by a  real parameter. This
parameter is  chosen using  a probability  distribution $Q_{x,y}$ that
depends on the initial position $(x,y)$ of the node.
 Then the randomly obtained curve is
shifted so that its startpoint coincides with the random
initial position of the node and then the node will move along this
random trajectory.

Let $d(x,y)$ be a nonnegative  real valued function over $D\times  D$,
with the  only restriction  that $d(x,x)=0$  holds for  any $x$.  This
function  is  intended  to  measure  ``radio  distance"  in  $D$.  The
assumption is that whenever $d(x,y)$  is small enough, then two  nodes
positioned at $x$ and $y$ can receive each others' transmissions.  The
function  $d(x,y)$,  however,  does  not  have  to  satisfy  the usual
distance   axioms,   it   may   reflect   complex   radio  propagation
characteristics, such as  expected attenuation and  fading,
it  may  account  for  the heterogeneity of the
terrain, for  propagation obstacles  etc. We  may also  include random
effects,  making  $d(x,y)$  a  random  variable,  reflecting   special
conditions of interest, such  as the random presence  of eavesdroppers
that can trigger  the inhibition of  certain links.
We  assume, however,  that if
there is randomness in $d(x,y)$,  then it is independent of  the other
random variables in the model.

Let $t_n$ and $r_n$ be further parameters that may also depend on  the
number  $n$  of  nodes.  We  now  define  the links of the network, as
follows.   Consider   two   nodes   with   initial   position  vectors
$X_1(0),X_2(0)$,  respectively.  As  they  move  along  their   random
trajectories,  their  positions  at  time  $t$  is denoted by $X_1(t),
X_2(t)$, respectively.  The two  nodes are  considered connected  by a
link, if there is a closed subinterval of length at least $t_n$ within
the time horizon $[0,T_n]$, such that $d(X_1(t),X_2(t))\leq r_n$ holds
for every time $t$  within the subinterval\footnote{The motivation  is
that the nodes should be within range at least for the time of sending
a packet.}, with the possibly complicated radio distance.

Now the question is this: for given $P$, $D$, $\cal C$, $Q_{x,y}$ and
$d(x,y)$,
and for the described way of dependent node genaration, can we somehow
choose the model parameters $k, d_0, v_0, T_n, t_n$ and $r_n$, such
that
the arising random graph is asymptotically almost surely connected,
while the expected average degree in the graph remains bounded?

We believe that it would be rather hard to answer such a question with
a direct analysis for arbitrary complex choices of $P$, $D$, $\cal  C$
$Q_{x,y}$ and $d(x,y)$. On the other hand, with our general results it
becomes quite straightforward, showing the strength of the results.

Let us choose the model domain $S$ as a 3-dimensional phase space,  in
which each  node is  represented by  a point  such that  the first two
coordinates describe  the intial  position of  the node  and the  last
coordinate encodes which  random trajectory was  chosen from $\cal  C$
for the node. Let $X_1,X_2,\ldots$ be the representations of the nodes
in this phase space.

We  can  now  check  that,  for  any  $n$,  the  joint distribution of
$X_1,\ldots,X_n$ is invariant to re-indexing them. The reason is  that
both the initial positions and the trajectory choices are generated by
processes in which  the indices do  not play any  role. Therefore, the
model  is  {\em  name  invariant}.  Interestingly,  this  remains true
despite having  a lot  of dependencies  among the  nodes: the  initial
positions of different nodes are not independent (due to  clustering),
and the  trajectory of  a given  node is  also not  independent of its
initial position, as it is drawn from a probability distribution  that
may  depend  on  the  location.  Through  this,  the  trajectories and
initial positions of different nodes also become dependent, making
their whole movement dependent. Yet, the model is still name
invariant.

Let us now consider the links. As defined above, two
nodes are considered connected if during their movement over the  time
horizon $[0,T_n]$ there is a  subinterval of time, of length  at least
$t_n$,  such  that  they  remain  within  ``radio distance" $\leq r_n$
during the  entire subinterval.  The radio  distance, however,  may be
very different from the Euclidean distance, it may be described by  an
arbitrary   function   that   may   account  for  complex  propagation
characteristics, attenuation, obstacles, and it may also contain
independent randomness.

Given some possibly complicated  radio distance $d(x,y)$ and  the node
generation and movement process with possibly complex trajectories, it
may not be easy to compute whether a link actually exists between  two
nodes according to the above definition. On the other hand, for us  it
is enough to note that once the phase space representations $X_i, X_j$
of any two  nodes are given,  plus the realization  of the independent
randomness of  the distance,  they together  determine whether  a link
exists between the two  nodes or not. The  reason is that the  initial
positions   and   the   trajectories,   given   in   the  phase  space
representation, fully determine the  movement of the nodes.  Once this
is known, it determines, along with the realization of the independent
randomness of the  distance function, whether  the link definition  is
satisfied, i.e.,  if there  is a  subinterval of  length $\geq t_n$ in
$[0,T_n]$, such that the nodes  stay within radio distance $\leq  r_n$
during the entire subinterval. To actually compute it may not be  easy
for a sophisticated case, but for our purposes it enough to know  that
it is {\em determined} by the listed factors, without knowing anything
about the other nodes. This implies that the model is {\em local}.

Thus, we have established that,
for any choice  of the  parameters,
 the problem can be described by a
model that is in ${\bf GEOM}(loc, inv)$.
Then this model cannot be in
${\bf BD}\,\wedge \,{\bf CONN}$, since we know from Theorem~\ref{thm3}
that
${\bf BD}\,\wedge \,\beta_n{\bf {\rm -}CONN}\,\wedge\,
{\bf GEOM}(loc,inv)=\emptyset$ holds for any choice of
$\beta_n\rightarrow
1$, including $\beta_n\equiv 1$. Thus, in our example it is impossible
to keep the expected average degree bounded and achieving
asymptotically almost sure
connectivity at the same time. With this we could cut through a lot of
complexity that would otherwise arise with the direct analysis of the
specific model.

\section{Conclusion and Open Problems}
\label{concl}

Our research has been motivated by the fact that many different random
graph constructions  are used  to model  large real  life networks,  but
often it is unclear how the strength of the different models compare
to each other, e.g., when will a certain model  property
imply another.  We have  set up  a general  framework to compare the
strength of  various random  graph model  classes, and  presented some
results about the equality, inequality and proper containment of these
classes. 

There are many research issues, however,  
that remain open. Let us  mention some examples that
seem interesting. They could lead to  nontrivial representation theorems for 
various model classes, and could clarify the relative strength of these classes.

\begin{description}

\item[Open problem  1.] One  can easily  see from  the definition that
${\bf GEOM}(loc, free)\preceq {\bf IDE}$. That is, in local  geometric
models  with  independent  node  variables  the  disjoint  edges   are
independent. Is the converse true,  i.e., can we represent any  ${\cal
M}\in {\bf  IDE}$ by  a local  geometric model  with independent  node
variables?

\item[Open problem 2.] Is it true that in every local and name
invariant geometric model the edges are positively correlated? In
other words, does
${\bf GEOM}(loc, inv)\preceq {\bf POS}$ hold? Or does at least
${\bf GEOM}(loc, free)\preceq {\bf POS}$ hold? Or else, what additional condition 
should be imposed to imply positive edge correlations?

\item[Open problem 3.] Is it true that
${\bf POS} \preceq{\bf GEOM}(loc, inv)$? If not, what
restrictions need to be added to $\bf POS$ to make it true?

\item[Open problem 4.] It is not hard to show via small examples that
$\bf IDE$ and $\bf POS$ are incomparable, that is, neither ${\bf
IDE}\preceq {\bf POS}$ nor ${\bf POS}\preceq {\bf IDE}$ hold.
Can the class ${\bf IDE}\wedge{\bf POS}$ be characterized in a
nontrivial way? How does it relate to ${\bf GEOM}(loc, inv)$?

\end{description}

\medskip\medskip
\medskip\medskip

\subsubsection*{Acknowledgment}

The authors gratefully acknowledge the support of
NSF Grant CNS-1018760.

\bibliographystyle{plain}
\bibliography{}

\end{document}